\documentclass[%
 aip, apl,
 amsmath,amssymb,
 reprint,%
 twocolumn,
]{revtex4-1}

\usepackage{amsmath,amssymb}
\usepackage{lipsum}
\usepackage{bm}
\usepackage{graphicx}
\usepackage{graphics}
\usepackage{amsmath}
\usepackage{array}
\usepackage[caption=false]{subfig}
\usepackage{xcolor}
\usepackage{subfig}
\usepackage{hyperref}
\usepackage[capitalise]{cleveref}
\usepackage{textcomp}

\usepackage{epstopdf}
\usepackage{cases}
\usepackage{amssymb}
\usepackage{multirow}
\usepackage{siunitx}
\usepackage{cases}
\usepackage{tabularx}
\usepackage[top=1.1in, bottom=1.1in, left=1.0in, right=1.0in]{geometry}
\usepackage[utf8]{inputenc}
\usepackage{tikz}
\usepackage{comment}
\usepackage{lineno,hyperref}
\modulolinenumbers[5]
\usepackage{booktabs}
\usepackage{diagbox}
\usepackage{graphicx}
\usepackage{dcolumn}
\usepackage{bm}

\usepackage{graphicx}
\usepackage{dcolumn}
\usepackage{bm}

\usepackage[utf8]{inputenc}
\usepackage[T1]{fontenc}
\usepackage{mathptmx}
\usepackage{etoolbox}

\makeatletter
\def\@email#1#2{%
 \endgroup
 \patchcmd{\titleblock@produce}
  {\frontmatter@RRAPformat}
  {\frontmatter@RRAPformat{\produce@RRAP{*#1\href{mailto:#2}{#2}}}\frontmatter@RRAPformat}
  {}{}
}%
\makeatother
\begin{document}
\preprint{AIP/APL}

\title{Transient hydrodynamic phonon transport in two-dimensional disk geometry}
\author{Chuang Zhang}
\email{zhangc520@hdu.edu.cn}
\affiliation{Department of Physics, School of Sciences, Hangzhou Dianzi University, Hangzhou 310018, China}
\author{Lei Wu}
\affiliation{Department of Mechanics and Aerospace Engineering, Southern University of Science and Technology, Shenzhen 518055, China}
\date{\today}

\begin{abstract}

Transient cooling phenomenon in two-dimensional materials is studied based on the phonon Boltzmann transport equation (BTE).
Using a heating laser pulse to heat the two-dimensional disk geometry under the environment temperature, after the heating laser is removed, our results show that the transient temperature could be lower than the environment temperature and this transient cooling phenomenon could only appear in the phonon hydrodynamic regime with sufficient normal-process and insufficient resistive-process.
In addition, the possibility of this phenomenon observed by experiments is theoretically discussed by using the experimental parameters as the input of the phonon BTE.
Results show that this transient cooling phenomenon could obviously appear in a single-layer suspended graphene disk with diameter $7~\mu$m in the temperature range of $50-150$ K.
The present work could provide theoretical guidance for the future experimental proof of hydrodynamic phonon transport in two-dimensional materials.

\end{abstract}


\maketitle

\section{INTRODUCTION}

Hydrodynamics is a kind of macroscopic phenomenon~\cite{prandtl_fluid,Gurzhi_1968,beck1974,Lucas_2018review,hydrodynamics_review_2022}, resulting from the frequent interactions between microscopic (quasi-)particles and satisfying the conservation principles of physical quantities, such as mass, (quasi-)momentum and energy.
Compared to the ubiquitous fluid hydrodynamics phenomena in our daily life~\cite{prandtl_fluid}, the observation of hydrodynamics behaviors in solid materials is much more difficult because the momentum is usually not conserved during phonon scattering processes~\cite{leesangyeopch1,chen_non-fourier_2021,hydrodynamics_review_2022}.

In order to observe the phonon hydrodynamics phenomena in solid materials, the momentum of the thermal system has to be highly conserved, which requires that the momentum conserved N-process should be much sufficient and meanwhile the R-process should be insufficient~\cite{PhysRev_GK,PhysRev.148.766,PhysRev.131.2013,sussmann1963,WangMr15application}.
In the early days, the phonon hydrodynamics phenomenon could only be measured in a few three-dimensional materials at sufficiently low temperatures (near $10$ K)~\cite{PhysRevLett.16.789,PhysRevLett_secondNaF,PhysRevLett_ssNaf,PhysRevLett.28.1461,PhysRevLett.99.265502,RevModPhysJoseph89}, which significantly limited its further development and research.
Fortunately, this dilemma was broken in the recent seven years~\cite{leesangyeopch1,chen_non-fourier_2021,yu_perspective_2021,LiShi_reexamination_hydrodynamic_JAP2022}.
In $2015$, researchers found that the N-process in suspended graphene is much more sufficient than the R-process, and the phonon hydrodynamics phenomena could be predicted theoretically at $100$ K~\cite{lee_hydrodynamic_2015,cepellotti_phonon_2015}.
These seminal work sets off an upsurge of theoretical and experimental research on phonon hydrodynamics phenomena in low-dimensional or multi-layer materials, such as the drifting second sound~\cite{lee2017,luo2019,XU2021127402,PhysRevB.105.165423}, heat vortices~\cite{shang_heat_2020} and phonon Poiseuille flow with parabolic distribution of the heat flux~\cite{Knudsenminimum_phonon78,li2018a,wangmr17callaway}.
The drifting second sound, more specifically, the temperature wave signal was measured experimentally by the transient thermal grating method in high-quality graphite samples at $100-200$ K~\cite{huberman_observation_2019,ding_observation_2022}.
Until now, the parabolic distribution of heat flux or the heat vortices has never been measured directly in experiments.
The Knudsen minimum phenomenon and temperature dependent thermal conductivity which increases faster than that in the ballistic limit~\cite{PhysRevLett_Strontium_Titanate,machida2020,machida2018,huang2023observation}, are usually regarded as the indirect macroscopic evidences of the phonon Poiseuille flow.
Above thermal behaviors have usually been regarded as the macroscopic signatures of the hydrodynamic phonon transport with sufficient N-process in previous studies~\cite{huberman_observation_2019,ding_observation_2022,leesangyeopch1}.

However, some studies have also reported that the temperature wave phenomenon could appear even if the N-process is insufficient~\cite{PhysRevB.101.075303,beardo_observation_2021,PhysRevB.104.245424,heatwaves_2022chuang}.
According to Hardy's work~\cite{PhysRevB_SECOND_SOUND}, when all phonons have similar relaxation time and there are external time-dependent heat sources whose heating period is comparable to the phonon relaxation time, the driftless second sound could appear.
And the drifting and driftless second sound have the same propagation speed at the macroscopic level when the linear phonon dispersion is accounted.
Based on first-principle and linearized phonon BTE, Cepellotti \textit{et al.} found that the driftless second sound can appear in two-dimensional materials even under the normal condition~\cite{cepellotti_phonon_2015}.
Recently, the ``high frequency'' second sound was measured by the frequency-domain thermoreflectance experiments in room temperature Germanium with insufficient N-process~\cite{beardo_observation_2021} when the heating frequency is high enough.
On the other hand, when the N-process is ignored, the heat vortices~\cite{ZHANG20191366,zhang_heat_2021} or the parabolic distribution of heat flux~\cite{PhysRevApplied.11.034003,WangMr15application} can also be predicted as long as the R-process is insufficient.
In a word, when the temperature wave, heat vortices or the parabolic distribution of heat flux is observed at the macroscopic/experimental level, the sufficient N-process is not necessarily required or the phonon transport may not be in the hydrodynamic regime.

Hence, if we want to experimentally prove the existence and uniqueness of hydrodynamic phonon transport, one of the best way might be to find a macroscopic phenomenon that only appears when the N-process is very sufficient and meanwhile the R-process is very insufficient.
In homogeneous 2D and 3D hotspot systems, the steady temperature distributions in the radial direction are stratified in different phonon transport regimes~\cite{chuang2021graded}, which may be another candidate.
For unsteady problem, a novel transient heat conduction phenomenon was reported based on frequency-independent BTE~\cite{zhang_transient_2021}, namely, the transient temperature could be lower than the lowest value of initial temperature when the external heat source is removed.
Furthermore, this transient cooling phenomenon could appear only when the N-process dominates heat conduction.
It was also measured by picosecond laser irradiation in a three-dimensional highly oriented pyrolytic graphite in the temperature range of $80-120$ K~\cite{PhysRevLett.127.085901}.
However, both two studies mainly focused on three-dimensional materials~\cite{zhang_transient_2021,PhysRevLett.127.085901}.

Therefore, our main goal in the present work is to figure out whether this transient cooling phenomenon could appear in two-dimensional materials, and whether it only appears in the hydrodynamic regime.
The rest of the paper is organized as follows.
The physical problem is described in Sec.~\ref{sec:Problemsettings} and the phonon BTE is introduced in Sec.~\ref{sec:PhononBTE}.
Results and discussions are shown in Sec.~\ref{sec:discussions}.
Finally, conclusions are made in Sec.~\ref{sec:conclusion}.

\section{Problem description}
\label{sec:Problemsettings}

\begin{figure}[htb]
 \centering
 \includegraphics[scale=1.0,clip=true]{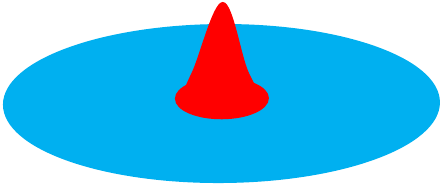}
 \caption{Schematic of the 2D disk geometry with a heat source at the center.}
 \label{problemdescription}
\end{figure}

As shown in~\cref{problemdescription}, in a two-dimensional disk geometry with diameter $L$, the initial environment temperature is $T_0$.
The outer of the disk is the heat sink with fixed temperature $T_0$.
When $t>0$, a heating pulsed Gaussian laser beam with radius $r_{pump}$ is added at the center and it continues to heat the system for a while.
The temperature rise is controlled much small.
When $t>t_h$, the heating pulse is removed and the temperature will dissipate with time gradually, where $t_h$ is the heating time.

The transient heat conduction process can be described by
\begin{align}
\frac{\partial E}{\partial t }+  \nabla \cdot  \bm{q} &= \dot{S} (\bm{x}, t),  \label{eq:firstlaw}
\end{align}
where $E$, $\bm{q}$ and $\dot{S} $ are the local energy, heat flux and the external heat source, respectively.
All of them depend on the spatial position $\bm{x}$ and time $t$.
The heating pulsed Gaussian laser beam is
\begin{align}
\begin{split}
\dot{S}= \left \{
\begin{array}{ll}
A \exp \left( - \frac{ |\bm{x}- \bm{x}_c |^2}{r_{pump} ^2 }  \right),                    & 0<t<t_h\\
0,   &  t \geq t_h
\end{array}
\right.
\end{split}
\end{align}
where $\bm{x}_c$ and $A$ are the center and peak of the heating pulse, respectively.
In practical pump-probe experiments~\cite{xu_raman-based_2020,fan_dual-wavelength_2019,beardo_observation_2021}, the measured temperature is usually a Gaussian average over the spatial range of the probe pulse.
Besides, the probe time is much shorter than the temperature decay time so that we ignore the average of the sampling time.
In this work, we mainly focus on the temperature variance at the disk center so that the probe temperature is
\begin{align}
T_{probe} (t) =\frac{   \int  T(\bm{x} , t ) \exp \left(  - \frac{ |\bm{x}- \bm{x}_c |^2}{r_{probe} ^2 }  \right)    d\bm{x}   }{ \int  \exp \left( - \frac{ |\bm{x}- \bm{x}_c |^2}{r_{probe} ^2 }  \right)  d\bm{x}    },
\label{eq:Tprobe}
\end{align}
where $r_{probe}$ is the laser spot radius of the probe pulse and $T$ is the local temperature.
In addition, we set
\begin{align}
T_h =  T_{probe} (t_h).
\end{align}
Obviously, $T_{probe} < T_h$ when the heating pulse is removed.

\section{Phonon BTE}
\label{sec:PhononBTE}

The key of Eq.~\eqref{eq:firstlaw} is the constitutive relation between the heat flux and temperature.
Unfortunately, the Fourier's law breaks down at the micro/nano scale.
In order to correctly describe the multiscale transient heat conduction process, the phonon Boltzmann transport equation (BTE) under the Callaway approximation is used~\cite{PhysRev_callaway,wangmr17callaway,luo2019,shang_heat_2020,nanoletterchengang_2018,heatwaves_2022chuang,zhang_transient_2021,nie2020thermal,WangMr15application},
\begin{align}
\frac{\partial e}{\partial t }+ v_g \bm{s} \cdot \nabla_{\bm{x}} e  &= \frac{e^{eq}_{R} -e}{\tau_{R}} + \frac{e^{eq}_{N}-e}{\tau_{N}}  + \frac{ \dot{s} }{2 \pi} ,  \label{eq:BTE}
\end{align}
where $e$ is the phonon distribution function of energy density, $v_g$ is the group velocity and $\bm{s}$ is the unit directional vector in two-dimensional solid angle space.
The N-process conserves both energy and momentum, while the R-process only conserves energy.
$e^{eq}_{R}$ and $e^{eq}_{N}$ are the associated phonon equilibrium distribution functions for R- and N- processes, respectively.
$\tau_R$ and $\tau_N$ are the relaxation times for R- and N- processes, respectively.
$\dot{s}=G \dot{S} $ is the spectral volumetric heat generation~\cite{APLnonthermal2020,PhysRevB.97.014307}, and $G$ is the associated weight satisfying
\begin{align}
\sum_p \int  G  d\omega =  1 ,
\end{align}
where the integral is conducted in the frequency space $d\omega$.
In general, $G$ is much complex and related to the actual physical nature of heat source~\cite{APLnonthermal2020,PhysRevB.93.125432}.

The thermalizing boundary conditions are used for the heat sink and all phonons emitted from the heat sink are the equilibrium distribution function $e^{eq}_R (T_0)$.
The details of phonon BTE can be found in Appendix~\ref{sec:SMBTE}.
The discrete unified gas kinetic scheme~\cite{guo_progress_DUGKS,luo2019} is used to numerically solve the frequency-independent phonon BTE, and the implicit kinetic scheme is used to solve the frequency-dependent phonon BTE, which have been validated in our previous studies~\cite{heatwaves_2022chuang,zhang_transient_2021}.
The numerical discretizations of the whole phase space are shown in Appendix~\ref{sec:SMsolver}.

\section{Results and Discussions}
\label{sec:discussions}

The transient heat conduction in 2D disk geometry with a fixed system size is simulated, and the Debye linear dispersion approximation and gray model are used~\cite{nie2020thermal,zhang_transient_2021,shang_heat_2020,WangMr15application}.
The heat conduction is not limited by specific materials properties and all physical variables are dimensionless.
We fixed $L=1$, $r_{pump}=r_{probe}=0.1$, $v_g=1$ and specific heat $C=1$.

\begin{figure}[htb]
\centering
\subfloat[Ultra short heating time $v_g t_h/L=0.01$]{\includegraphics[scale=0.45,clip=true]{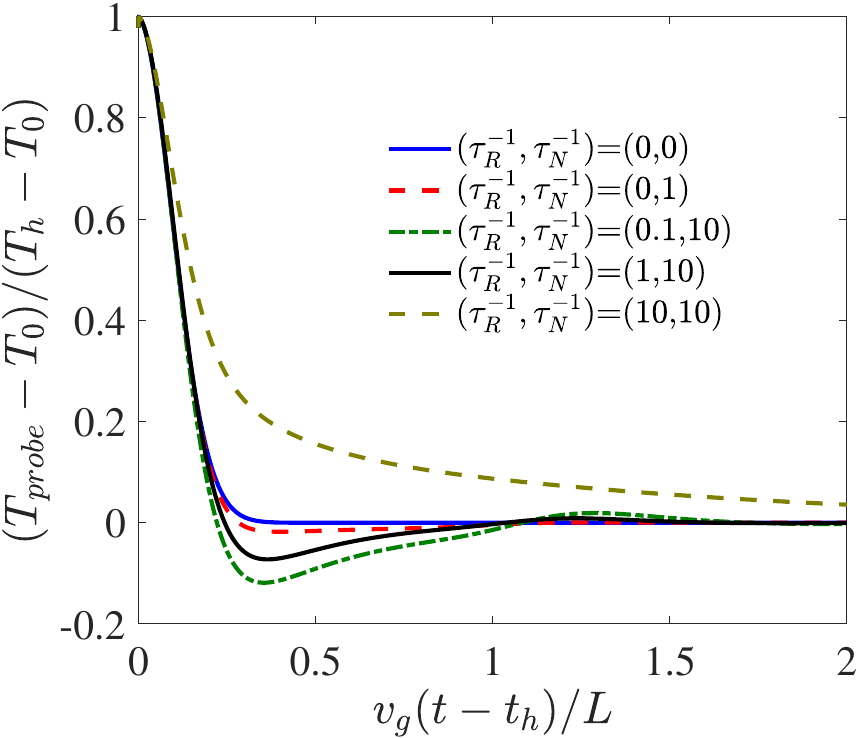} }  \\
\subfloat[Ultra long heating time $v_g t_h/L =5.0$]{\includegraphics[scale=0.45,clip=true]{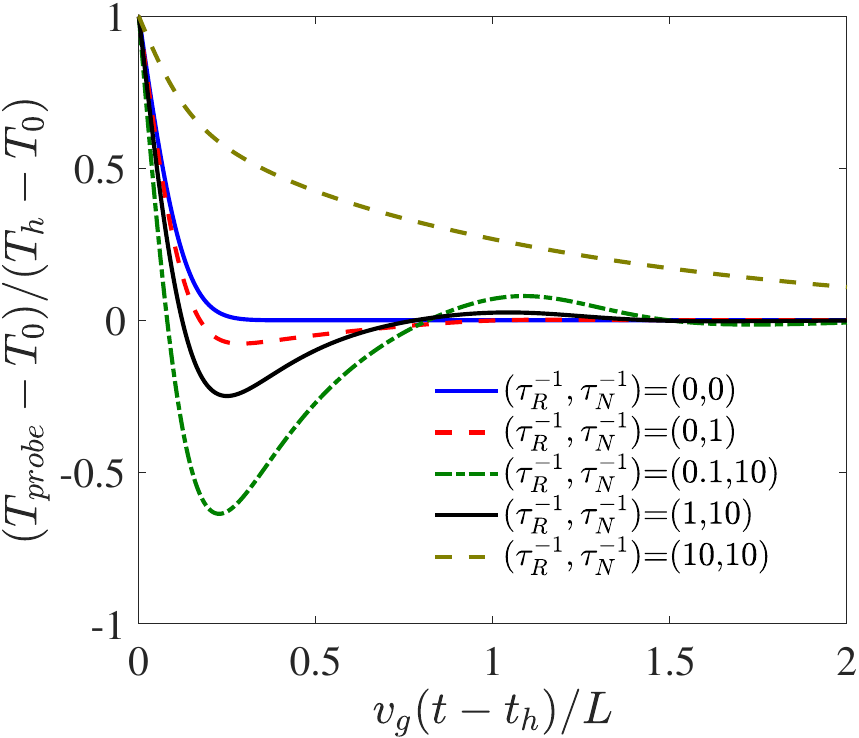} }
\caption{The evolution of temperature~\eqref{eq:Tprobe} in 2D disk geometry.  }
\label{graymodel}
\end{figure}

The evolution of the probe temperature~\eqref{eq:Tprobe} is shown in~\cref{graymodel} after the heat source is removed.
It can be found that $(T_{probe}-T_0)/(T_h -T_0 ) \geq 0$ when both the N- and R- processes are rare.
However, when the N-process is more sufficient than the R-process, the transient temperature could be smaller than the environment temperature when $v_g(t-t_h)/L <0.45$.
Furthermore, this phenomenon becomes more obvious with the increase of N-process.
When the R-process increases, this phenomenon becomes weak and finally disappears.
In addition, based on the results in~\cref{graymodel}(a)(b) with different heating time $t_h$, it can be found that the appearance of this phenomenon in the hydrodynamic regime is slightly affected by the heating time $t_h$ of the laser beam.

\begin{figure}[htb]
\centering
\includegraphics[scale=0.6,clip=true]{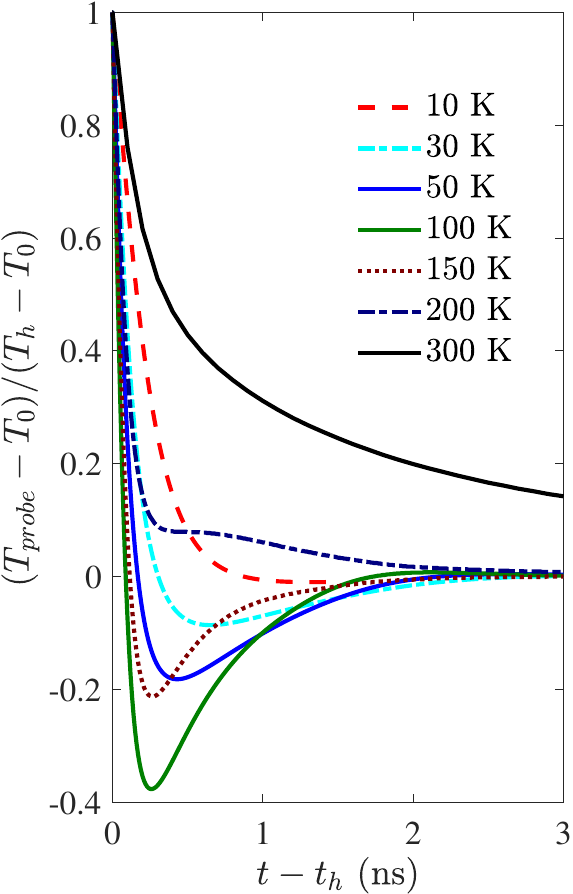}~~
\caption{The evolution of temperature~\eqref{eq:Tprobe} in graphene disk under different environment temperature, where $t_h =20$ ns.  }
\label{transient2D}
\end{figure}
The underlying physical mechanisms are discussed below.
In the ballistic regime, phonons transport with rare scattering so that
\begin{align}
e(\bm{x},\bm{s},t+ \delta t) = e(\bm{x} -v_g \bm{s} \delta t,\bm{s},t ), \quad t \geq t_h,
\end{align}
where $\delta t >0$ is an arbitrary time increment.
The phonon distribution function inside the thermal system is totally controlled by the boundary conditions or the phonon distribution function at $t=t_h$.
Note that the local temperature is a sum total of all phonon distribution functions with different directions (Eq.(S18)), so that it is not smaller than the environment temperature.
When the phonon N- or R- process scattering increases, the phonon transport can be approximately described by a hyperbolic heat conduction equation~\cite{PhysRevX.10.011019,beardo_observation_2021,cepellotti_phonon_2015,zhang_transient_2021,shang_heat_2020},
\begin{align}
c_1 \frac{ \partial^2 T }{ \partial t^2 } + c_2  \frac{ \partial T }{ \partial t } + c_3 \nabla_{\bm{x}}^2 T = \dot{S},
\label{eq:wavehyperbolic}
\end{align}
where $c_1$, $c_2$, $c_3$ are coefficients.
When the N-process dominates, the heat wave term in Eq.~\eqref{eq:wavehyperbolic} plays an important role on the transient heat conduction~\cite{PhysRevX.10.011019,shang_heat_2020,cepellotti_phonon_2015}.
As long as the heat conduction in thermal systems could be described by a hyperbolic equation and the wave term dominates, the temperature could fluctuate around the environment temperature within suitable initial and boundary conditions~\cite{zhang_transient_2021}.
That's a feature of the wave equation.
It is not limited by the system size, dimensions and materials properties.
When the R-process increases, the diffusion term dominates heat conduction and the temperature fluctuation disappears.

In a word, the negative dimensionless temperature $(T_{probe}-T_0)/(T_h -T_0 ) $ could appear in two-dimensional materials.
More importantly, different from the heat wave which can appear in both the ballistic and hydrodynamic regimes~\cite{heatwaves_2022chuang,PhysRevB.104.245424,beardo_observation_2021}, this transient cooling phenomenon could only appear in the phonon hydrodynamic regime.
Therefore, this transient cooling phenomenon could be a unique macroscopic signature of the hydrodynamic phonon transport.

To make our conclusion more convincing, the heat conduction in a single-layer suspended graphene disk is studied due to its representativeness in two-dimensional materials~\cite{cepellotti_phonon_2015,lee_hydrodynamic_2015,RevModPhys.90.041002}, and the possibility of future experimental measurements is discussed.
The experimental parameters of dual-wavelength flash Raman mapping method~\cite{fan_dual-wavelength_2019,xu_raman-based_2020} provided by Dr. Aoran Fan are used: $L=7~\mu$m, $r_{pump}$=259 nm, $r_{probe}=282$ nm and $t_h=20$ ns.
Note that $t_h=20$ ns is long enough for sufficient phonon scattering and reaching steady-state.
The phonon dispersion and polarization of graphene are calculated by the Vienna \textit{Ab initio} Simulation Package (VASP) combined with phonopy.
Details can be found in our previous paper~\cite{chuang2021graded}.
The physical nature of the photothermal conversion including photon-electron-phonon coupling is quite complicated~\cite{PhysRevB.93.125432,MTP_review_nuo_2021,ChenG05Oxford}, so that here we assume that all phonons reach thermal equilibrium after absorbing the energy coming from heating laser beam~\cite{APLnonthermal2020}.
The weight depends on the specific heat,
\begin{align}
G=\frac{C}{ \sum_p \int C d\omega   }.
\end{align}

After the heat source is removed, the evolution of the probe temperature~\eqref{eq:Tprobe} under different environment temperature is shown in~\cref{transient2D}.
In room temperature, the heat dissipates gradually with time until reaching environment temperature.
When the environment temperature decreases, the negative normalized temperature $(T_{probe}-T_0)/(T_h -T_0 ) $  is probed.
Namely, the transient temperature could be smaller than the environment temperature.
This phenomenon is quite obvious in the temperature range of $50-150$ K.
When the environment temperature further decreases, it can be observed that this phenomenon becomes weaker and weaker, and finally fades away at $T_0 =10$ K.

As reported in the previous studies~\cite{lee_hydrodynamic_2015,cepellotti_phonon_2015,luo2019}, when the environment temperature decreases from high temperature to extremely low temperature, the phonon transport in graphene could go through the diffusive, hydrodynamic and ballistic regimes in turn.
In other words, the negative dimensionless temperature phenomenon could appear in the graphene disk and only appear in the hydrodynamic regime.

\section{Conclusion}
\label{sec:conclusion}

In summary, a novel transient cooling phenomenon in two-dimensional materials is studied based on the phonon BTE under the Callaway model.
Results show that the transient temperature in two-dimensional disk geometry could be lower than the environment temperature and this transient cooling phenomenon could only appear in the hydrodynamic regime.
Furthermore, by using the experimental parameters as the input of the phonon BTE, we found that this phenomenon could obviously appear in a single-layer suspended graphene disk with diameter $7~\mu$m in the temperature range of $50-150$ K.
This work provides theoretical guidance for experimentally proving the existence and uniqueness of hydrodynamic phonon transport in two-dimensional materials in the future.

%
%
%

\section*{Acknowledgments}

This work is supported by National Natural Science Foundation of China (12147122).
The authors acknowledge Dr. Aoran Fan, Rulei Guo and Yufeng Zhang for their enthusiastic guidance and help in the Raman/pump-probe experimental details and parameters.

\appendix

\section{Phonon Boltzmann Transport Equation}
\label{sec:SMBTE}

In order to correctly describe the multiscale transient heat conduction process in two-dimensional materials, the phonon Boltzmann transport equation (BTE) under the Callaway approximation is used~\cite{PhysRev_callaway,wangmr17callaway,luo2019,shang_heat_2020,nanoletterchengang_2018,heatwaves_2022chuang,zhang_transient_2021,nie2020thermal,WangMr15application},
\begin{align}
\frac{\partial e}{\partial t }+ v_g \bm{s} \cdot \nabla_{\bm{x}} e  &= \frac{e^{eq}_{R} -e}{\tau_{R}} + \frac{e^{eq}_{N}-e}{\tau_{N}}  + \frac{ \dot{s} }{2 \pi} ,  \label{eq:BTE}  \\
e & = \hbar \omega D (f-f_{BE}(T_0)) /(2 \pi),  \\
e^{eq}_R & = \hbar \omega D ( f^{eq}_R-f_{BE}(T_0)) /(2 \pi),  \\
e^{eq}_N & = \hbar \omega D ( f^{eq}_N-f_{BE}(T_0)) /(2 \pi),
\end{align}
where $f=f(\bm{x},\bm{K}, \omega, p ,t)$ is the phonon distribution function, depending on spatial position $\bm{x}$, wave vector $\bm{K}$, phonon angular frequency $\omega$, polarization $p$ and time $t$.
The two-dimensional wave vector space is assumed isotropic.
$v_g$ is the group velocity and $\bm{s}$ is the unit directional vector in two-dimensional solid angle space satisfying $v_g \bm{s} = \nabla_{\bm{K} } {\omega}$.
$e$ is the phonon distribution function of energy density, $\hbar$ is the Planck constant reduced by $2\pi$, $D=|\bm{K}|/\left( 2 \pi v_g \right)$ is the phonon density of state, $T$ is the temperature and $T_0$ is the reference temperature.
$\tau_R$ and $\tau_N$ are the relaxation times for R- and N- processes, respectively.
The N-process conserves both energy and momentum, while the R-process only conserves energy.
The R-process is a combination of all momentum destroying phonon scattering process except the boundary scattering based on the Mathiessen's rule~\cite{ChenG05Oxford}, including Umklapp, impurity, isotope process scattering.
$f^{eq}_{R}$ is the equilibrium distribution functions for R-processes, satisfying the Bose-Einstein distribution $f_{BE}$,
\begin{align}
 f^{eq}_{R}(T)   =  f_{BE} (T) =\frac{1}{ \exp \left( \frac{\hbar \omega }{k_B T} \right) -1 },
\end{align}
where $k_B$ is the Boltzmann constant.
$f^{eq}_{N}$ is the equilibrium distribution functions for N-processes,
\begin{align}
f^{eq}_{N}(T,\bm{u})= \frac{1}{ \exp \left( \frac{\hbar \omega - \hbar \bm{K} \cdot \bm{u} }{k_B T} \right) -1 }, \label{eq:feqnormal}
\end{align}
where $\bm{u}$ is the drift velocity.
$\dot{s}=G \dot{S} $ is the spectral volumetric heat generation~\cite{APLnonthermal2020,PhysRevB.97.014307}, and $G$ is the associated weight satisfying
\begin{align}
\sum_p \int  G  d\omega =  1 ,
\end{align}
where the integral is conducted in the frequency space $d\omega$.
In general, $G$ is much complex and related to the actual physical nature of heat source~\cite{APLnonthermal2020,PhysRevB.93.125432}.
When the phonon frequency $\omega$ and polarization $p$ are not considered, $G=1$.

Assuming that the temperature difference in the whole domain is much small, i.e., $|T- T_0| \ll T_0$, then the equilibrium distribution functions can be linearized,
\begin{align}
e^{eq}_{R}(T) &\approx C \frac{ T-T_0}{2 \pi } , \label{eq:feqR} \\
e^{eq}_{N}(T,\bm{u}) &\approx  C \frac{ T-T_0}{2 \pi }  +CT \frac{\bm{K} \cdot \bm{u} }{ 2 \pi \omega},\label{eq:feqN}
\end{align}
where $C $ is the mode specific heat at $T_0$,
\begin{equation}
C(\omega,p,T_0)= \hbar \omega D \left.\ \frac{  \partial{  f _{BE}}}{ \partial{T}} \right|_{T=T_0} .
\label{eq:specificheat}
\end{equation}
Due to the conservation principle of the phonon N- and R- scattering processes, we have
\begin{align}
0  &=  \sum_{p} \int \int  \frac{e^{eq}_{R}(T_R) -e}{\tau_{R}(T_0) }   d\Omega d\omega , \label{eq:Rconsertvation}  \\
0  &=  \sum_{p} \int \int    \frac{e^{eq}_{N} (T_N)-e}{\tau_{N}(T_0) }   d\Omega d\omega , \label{eq:Nconsertvation}  \\
0 &=  \sum_{p} \int \int  \frac{\bm{K} }{\omega } \frac{e^{eq}_{N} (T_N,\bm{u})-e}{\tau_{N}(T_0) }   d\Omega d\omega . \label{eq:NMconsertvation}
\end{align}
where $T_R$ and $T_N$ are local pseudotemperatures, and the integral is conducted in both the two-dimensional solid angle space $d\Omega$ and frequency space $d\omega$.
Based on above three equations, the macroscopic variables $T_R$, $T_N$ and $\bm{u}$ can be obtained,
\begin{align}
T_{R} & =T_{0}+ \left( \sum_{p}\int  \frac{\int e d\Omega }{\tau_R } d{\omega}   \right) \times \left(  \sum_{p}\int \frac{ C}{\tau_R }  d{\omega} \right)^{-1} , \\
T_{N} & =T_{0}+ \left( \sum_{p}\int  \frac{\int e d\Omega }{\tau_N } d{\omega}   \right) \times \left(  \sum_{p}\int \frac{ C}{\tau_N }  d{\omega} \right)^{-1} , \\
\bm{u}  &=  \frac{2}{T_{N}  \sum_{p} \int \frac{ |\bm{K}|^2  }{ \omega^2  } \frac{ C }{\tau_N } d{\omega}  } \sum_{p} \int \int  \frac{ \bm{K} }{\omega} \frac{ e }{ \tau_N}   d\Omega d\omega.
\end{align}

The local energy density, temperature and heat flux are obtained by taking the moments of the distribution function~\cite{Kubo1991statistical,ChenG05Oxford},
\begin{align}
E &= \sum_{p} \int \int e d\Omega d\omega,  \label{eq:energy} \\
T  &=T_0+ \frac{ \sum_{p} \int \int e d\Omega d\omega } { \sum_{p} \int C d\omega  },   \label{eq:T}  \\
\bm{q} &=  \sum_{p} \int \int \bm{v} e d\Omega d\omega ,
\label{eq:heatflux}
\end{align}
Note that the temperature cannot be well defined in non-equilibrium thermal systems~\cite{Kubo1991statistical}, so that the temperature in this study is more like the symbol of local energy density~\cite{Kubo1991statistical,ChenG05Oxford}.

\section{Numerical discretizations of the phase space}
\label{sec:SMsolver}

The spatial space is discretized with $201 \times 201$ or $401 \times 401$ uniform cells.
The first-order upwind scheme is used to deal with the spatial gradient of the distribution function in the ballistic regime, and the van Leer limiter is used in both the hydrodynamic and diffusive regimes.
The time step is $\Delta t= \text{CFL} \times \Delta x  / v_{\text{max}} $,
where $\Delta x$ is the minimum discretized cell size, $\text{CFL}$ is the Courant–Friedrichs–Lewy number and $v_{\text{max}} $ is the maximum group velocity.
In the two-dimensional solid angle space, $\bm{s}=\left( \cos \theta, \sin \theta  \right)$, where $\theta \in [0, 2 \pi]$ is the polar angle.
Due to symmetry, the $\theta \in [0,\pi]$ is discretized with the $N_{\theta}$-point Gauss-Legendre quadrature.
In graphene disk, $300$ discretized frequency bands are considered~\cite{heatwaves_2022chuang,chuang2021graded} and the mid-point rule is used for the numerical integration of the frequency space.
For all cases, $N_{\theta}=50$ and $\text{CFL}=0.04-0.4$.
The present discretizations of the phase space are enough for all phonon transport regimes according to our previous experience~\cite{shang_heat_2020,heatwaves_2022chuang,chuang2021graded,zhang_transient_2021}.

\bibliography{phonon}

\begin{thebibliography}{61}%
\makeatletter
\providecommand \@ifxundefined [1]{%
 \@ifx{#1\undefined}
}%
\providecommand \@ifnum [1]{%
 \ifnum #1\expandafter \@firstoftwo
 \else \expandafter \@secondoftwo
 \fi
}%
\providecommand \@ifx [1]{%
 \ifx #1\expandafter \@firstoftwo
 \else \expandafter \@secondoftwo
 \fi
}%
\providecommand \natexlab [1]{#1}%
\providecommand \enquote  [1]{``#1''}%
\providecommand \bibnamefont  [1]{#1}%
\providecommand \bibfnamefont [1]{#1}%
\providecommand \citenamefont [1]{#1}%
\providecommand \href@noop [0]{\@secondoftwo}%
\providecommand \href [0]{\begingroup \@sanitize@url \@href}%
\providecommand \@href[1]{\@@startlink{#1}\@@href}%
\providecommand \@@href[1]{\endgroup#1\@@endlink}%
\providecommand \@sanitize@url [0]{\catcode `\\12\catcode `\$12\catcode
  `\&12\catcode `\#12\catcode `\^12\catcode `\_12\catcode `\%12\relax}%
\providecommand \@@startlink[1]{}%
\providecommand \@@endlink[0]{}%
\providecommand \url  [0]{\begingroup\@sanitize@url \@url }%
\providecommand \@url [1]{\endgroup\@href {#1}{\urlprefix }}%
\providecommand \urlprefix  [0]{URL }%
\providecommand \Eprint [0]{\href }%
\providecommand \doibase [0]{http://dx.doi.org/}%
\providecommand \selectlanguage [0]{\@gobble}%
\providecommand \bibinfo  [0]{\@secondoftwo}%
\providecommand \bibfield  [0]{\@secondoftwo}%
\providecommand \translation [1]{[#1]}%
\providecommand \BibitemOpen [0]{}%
\providecommand \bibitemStop [0]{}%
\providecommand \bibitemNoStop [0]{.\EOS\space}%
\providecommand \EOS [0]{\spacefactor3000\relax}%
\providecommand \BibitemShut  [1]{\csname bibitem#1\endcsname}%
\let\auto@bib@innerbib\@empty
\bibitem [{\citenamefont {jr.}(2010)}]{prandtl_fluid}%
  \BibitemOpen
  \bibfield  {author} {\bibinfo {author} {\bibfnamefont {H.~O.}\ \bibnamefont
  {jr.}},\ }\href {\doibase 10.1007/978-1-4419-1564-1} {\emph {\bibinfo {title}
  {Prandtl-Essentials of Fluid Mechanics}}},\ \bibinfo {edition} {3rd}\ ed.,\
  \bibinfo {series} {Applied Mathematical Sciences}, Vol.\ \bibinfo {volume}
  {158}\ (\bibinfo  {publisher} {Springer-Verlag New York},\ \bibinfo {year}
  {2010})\BibitemShut {NoStop}%
\bibitem [{\citenamefont {Gurzhi}(1968)}]{Gurzhi_1968}%
  \BibitemOpen
  \bibfield  {author} {\bibinfo {author} {\bibfnamefont {R.~N.}\ \bibnamefont
  {Gurzhi}},\ }\href {\doibase 10.1070/pu1968v011n02abeh003815} {\bibfield
  {journal} {\bibinfo  {journal} {Sov. Phys.-Usp.}\ }\textbf {\bibinfo {volume}
  {11}},\ \bibinfo {pages} {255} (\bibinfo {year} {1968})}\BibitemShut
  {NoStop}%
\bibitem [{\citenamefont {Beck}, \citenamefont {Meier},\ and\ \citenamefont
  {Thellung}(1974)}]{beck1974}%
  \BibitemOpen
  \bibfield  {author} {\bibinfo {author} {\bibfnamefont {H.}~\bibnamefont
  {Beck}}, \bibinfo {author} {\bibfnamefont {P.~F.}\ \bibnamefont {Meier}}, \
  and\ \bibinfo {author} {\bibfnamefont {A.}~\bibnamefont {Thellung}},\ }\href
  {\doibase 10.1002/pssa.2210240102} {\bibfield  {journal} {\bibinfo  {journal}
  {Phys. Status Solidi A}\ }\textbf {\bibinfo {volume} {24}},\ \bibinfo {pages}
  {11} (\bibinfo {year} {1974})}\BibitemShut {NoStop}%
\bibitem [{\citenamefont {Lucas}\ and\ \citenamefont
  {Fong}(2018)}]{Lucas_2018review}%
  \BibitemOpen
  \bibfield  {author} {\bibinfo {author} {\bibfnamefont {A.}~\bibnamefont
  {Lucas}}\ and\ \bibinfo {author} {\bibfnamefont {K.~C.}\ \bibnamefont
  {Fong}},\ }\href {\doibase 10.1088/1361-648x/aaa274} {\bibfield  {journal}
  {\bibinfo  {journal} {J. Phys-Condens. Mat.}\ }\textbf {\bibinfo {volume}
  {30}},\ \bibinfo {pages} {053001} (\bibinfo {year} {2018})}\BibitemShut
  {NoStop}%
\bibitem [{\citenamefont {Ghosh}, \citenamefont {Kusiak},\ and\ \citenamefont
  {Battaglia}(2022)}]{hydrodynamics_review_2022}%
  \BibitemOpen
  \bibfield  {author} {\bibinfo {author} {\bibfnamefont {K.}~\bibnamefont
  {Ghosh}}, \bibinfo {author} {\bibfnamefont {A.}~\bibnamefont {Kusiak}}, \
  and\ \bibinfo {author} {\bibfnamefont {J.-L.}\ \bibnamefont {Battaglia}},\
  }\href {\doibase 10.1088/1361-648x/ac718a} {\bibfield  {journal} {\bibinfo
  {journal} {J. Phys-Condens. Mat.}\ }\textbf {\bibinfo {volume} {34}},\
  \bibinfo {pages} {323001} (\bibinfo {year} {2022})}\BibitemShut {NoStop}%
\bibitem [{\citenamefont {Lee}\ and\ \citenamefont
  {Li}(2020)}]{leesangyeopch1}%
  \BibitemOpen
  \bibfield  {author} {\bibinfo {author} {\bibfnamefont {S.}~\bibnamefont
  {Lee}}\ and\ \bibinfo {author} {\bibfnamefont {X.}~\bibnamefont {Li}},\ }in\
  \href {\doibase 10.1088/978-0-7503-1738-2ch1} {\emph {\bibinfo {booktitle}
  {Nanoscale Energy Transport}}},\ \bibinfo {series and number} {2053-2563}\
  (\bibinfo  {publisher} {IOP Publishing},\ \bibinfo {year} {2020})\ pp.\
  \bibinfo {pages} {1--1 to 1--26}\BibitemShut {NoStop}%
\bibitem [{\citenamefont {Chen}(2021)}]{chen_non-fourier_2021}%
  \BibitemOpen
  \bibfield  {author} {\bibinfo {author} {\bibfnamefont {G.}~\bibnamefont
  {Chen}},\ }\href {\doibase 10.1038/s42254-021-00334-1} {\bibfield  {journal}
  {\bibinfo  {journal} {Nat. Rev. Phys.}\ }\textbf {\bibinfo {volume} {3}},\
  \bibinfo {pages} {555} (\bibinfo {year} {2021})}\BibitemShut {NoStop}%
\bibitem [{\citenamefont {Guyer}\ and\ \citenamefont
  {Krumhansl}(1966{\natexlab{a}})}]{PhysRev_GK}%
  \BibitemOpen
  \bibfield  {author} {\bibinfo {author} {\bibfnamefont {R.~A.}\ \bibnamefont
  {Guyer}}\ and\ \bibinfo {author} {\bibfnamefont {J.~A.}\ \bibnamefont
  {Krumhansl}},\ }\href {\doibase 10.1103/PhysRev.148.778} {\bibfield
  {journal} {\bibinfo  {journal} {Phys. Rev.}\ }\textbf {\bibinfo {volume}
  {148}},\ \bibinfo {pages} {778} (\bibinfo {year}
  {1966}{\natexlab{a}})}\BibitemShut {NoStop}%
\bibitem [{\citenamefont {Guyer}\ and\ \citenamefont
  {Krumhansl}(1966{\natexlab{b}})}]{PhysRev.148.766}%
  \BibitemOpen
  \bibfield  {author} {\bibinfo {author} {\bibfnamefont {R.~A.}\ \bibnamefont
  {Guyer}}\ and\ \bibinfo {author} {\bibfnamefont {J.~A.}\ \bibnamefont
  {Krumhansl}},\ }\href {\doibase 10.1103/PhysRev.148.766} {\bibfield
  {journal} {\bibinfo  {journal} {Phys. Rev.}\ }\textbf {\bibinfo {volume}
  {148}},\ \bibinfo {pages} {766} (\bibinfo {year}
  {1966}{\natexlab{b}})}\BibitemShut {NoStop}%
\bibitem [{\citenamefont {Chester}(1963)}]{PhysRev.131.2013}%
  \BibitemOpen
  \bibfield  {author} {\bibinfo {author} {\bibfnamefont {M.}~\bibnamefont
  {Chester}},\ }\href {\doibase 10.1103/PhysRev.131.2013} {\bibfield  {journal}
  {\bibinfo  {journal} {Phys. Rev.}\ }\textbf {\bibinfo {volume} {131}},\
  \bibinfo {pages} {2013} (\bibinfo {year} {1963})}\BibitemShut {NoStop}%
\bibitem [{\citenamefont {Sussmann}\ and\ \citenamefont
  {Thellung}(1963)}]{sussmann1963}%
  \BibitemOpen
  \bibfield  {author} {\bibinfo {author} {\bibfnamefont {J.~A.}\ \bibnamefont
  {Sussmann}}\ and\ \bibinfo {author} {\bibfnamefont {A.}~\bibnamefont
  {Thellung}},\ }\href {\doibase 10.1088/0370-1328/81/6/318} {\bibfield
  {journal} {\bibinfo  {journal} {Proc. Phys. Soc.}\ }\textbf {\bibinfo
  {volume} {81}},\ \bibinfo {pages} {1122} (\bibinfo {year}
  {1963})}\BibitemShut {NoStop}%
\bibitem [{\citenamefont {Guo}\ and\ \citenamefont
  {Wang}(2015)}]{WangMr15application}%
  \BibitemOpen
  \bibfield  {author} {\bibinfo {author} {\bibfnamefont {Y.}~\bibnamefont
  {Guo}}\ and\ \bibinfo {author} {\bibfnamefont {M.}~\bibnamefont {Wang}},\
  }\href {\doibase 10.1016/j.physrep.2015.07.003} {\bibfield  {journal}
  {\bibinfo  {journal} {Phys. Rep.}\ }\textbf {\bibinfo {volume} {595}},\
  \bibinfo {pages} {1 } (\bibinfo {year} {2015})}\BibitemShut {NoStop}%
\bibitem [{\citenamefont {Ackerman}\ \emph {et~al.}(1966)\citenamefont
  {Ackerman}, \citenamefont {Bertman}, \citenamefont {Fairbank},\ and\
  \citenamefont {Guyer}}]{PhysRevLett.16.789}%
  \BibitemOpen
  \bibfield  {author} {\bibinfo {author} {\bibfnamefont {C.~C.}\ \bibnamefont
  {Ackerman}}, \bibinfo {author} {\bibfnamefont {B.}~\bibnamefont {Bertman}},
  \bibinfo {author} {\bibfnamefont {H.~A.}\ \bibnamefont {Fairbank}}, \ and\
  \bibinfo {author} {\bibfnamefont {R.~A.}\ \bibnamefont {Guyer}},\ }\href
  {\doibase 10.1103/PhysRevLett.16.789} {\bibfield  {journal} {\bibinfo
  {journal} {Phys. Rev. Lett.}\ }\textbf {\bibinfo {volume} {16}},\ \bibinfo
  {pages} {789} (\bibinfo {year} {1966})}\BibitemShut {NoStop}%
\bibitem [{\citenamefont {McNelly}\ \emph {et~al.}(1970)\citenamefont
  {McNelly}, \citenamefont {Rogers}, \citenamefont {Channin}, \citenamefont
  {Rollefson}, \citenamefont {Goubau}, \citenamefont {Schmidt}, \citenamefont
  {Krumhansl},\ and\ \citenamefont {Pohl}}]{PhysRevLett_secondNaF}%
  \BibitemOpen
  \bibfield  {author} {\bibinfo {author} {\bibfnamefont {T.~F.}\ \bibnamefont
  {McNelly}}, \bibinfo {author} {\bibfnamefont {S.~J.}\ \bibnamefont {Rogers}},
  \bibinfo {author} {\bibfnamefont {D.~J.}\ \bibnamefont {Channin}}, \bibinfo
  {author} {\bibfnamefont {R.~J.}\ \bibnamefont {Rollefson}}, \bibinfo {author}
  {\bibfnamefont {W.~M.}\ \bibnamefont {Goubau}}, \bibinfo {author}
  {\bibfnamefont {G.~E.}\ \bibnamefont {Schmidt}}, \bibinfo {author}
  {\bibfnamefont {J.~A.}\ \bibnamefont {Krumhansl}}, \ and\ \bibinfo {author}
  {\bibfnamefont {R.~O.}\ \bibnamefont {Pohl}},\ }\href {\doibase
  10.1103/PhysRevLett.24.100} {\bibfield  {journal} {\bibinfo  {journal} {Phys.
  Rev. Lett.}\ }\textbf {\bibinfo {volume} {24}},\ \bibinfo {pages} {100}
  (\bibinfo {year} {1970})}\BibitemShut {NoStop}%
\bibitem [{\citenamefont {Jackson}, \citenamefont {Walker},\ and\ \citenamefont
  {McNelly}(1970)}]{PhysRevLett_ssNaf}%
  \BibitemOpen
  \bibfield  {author} {\bibinfo {author} {\bibfnamefont {H.~E.}\ \bibnamefont
  {Jackson}}, \bibinfo {author} {\bibfnamefont {C.~T.}\ \bibnamefont {Walker}},
  \ and\ \bibinfo {author} {\bibfnamefont {T.~F.}\ \bibnamefont {McNelly}},\
  }\href {\doibase 10.1103/PhysRevLett.25.26} {\bibfield  {journal} {\bibinfo
  {journal} {Phys. Rev. Lett.}\ }\textbf {\bibinfo {volume} {25}},\ \bibinfo
  {pages} {26} (\bibinfo {year} {1970})}\BibitemShut {NoStop}%
\bibitem [{\citenamefont {Narayanamurti}\ and\ \citenamefont
  {Dynes}(1972)}]{PhysRevLett.28.1461}%
  \BibitemOpen
  \bibfield  {author} {\bibinfo {author} {\bibfnamefont {V.}~\bibnamefont
  {Narayanamurti}}\ and\ \bibinfo {author} {\bibfnamefont {R.~C.}\ \bibnamefont
  {Dynes}},\ }\href {\doibase 10.1103/PhysRevLett.28.1461} {\bibfield
  {journal} {\bibinfo  {journal} {Phys. Rev. Lett.}\ }\textbf {\bibinfo
  {volume} {28}},\ \bibinfo {pages} {1461} (\bibinfo {year}
  {1972})}\BibitemShut {NoStop}%
\bibitem [{\citenamefont {Koreeda}, \citenamefont {Takano},\ and\ \citenamefont
  {Saikan}(2007)}]{PhysRevLett.99.265502}%
  \BibitemOpen
  \bibfield  {author} {\bibinfo {author} {\bibfnamefont {A.}~\bibnamefont
  {Koreeda}}, \bibinfo {author} {\bibfnamefont {R.}~\bibnamefont {Takano}}, \
  and\ \bibinfo {author} {\bibfnamefont {S.}~\bibnamefont {Saikan}},\ }\href
  {\doibase 10.1103/PhysRevLett.99.265502} {\bibfield  {journal} {\bibinfo
  {journal} {Phys. Rev. Lett.}\ }\textbf {\bibinfo {volume} {99}},\ \bibinfo
  {pages} {265502} (\bibinfo {year} {2007})}\BibitemShut {NoStop}%
\bibitem [{\citenamefont {Joseph}\ and\ \citenamefont
  {Preziosi}(1989)}]{RevModPhysJoseph89}%
  \BibitemOpen
  \bibfield  {author} {\bibinfo {author} {\bibfnamefont {D.~D.}\ \bibnamefont
  {Joseph}}\ and\ \bibinfo {author} {\bibfnamefont {L.}~\bibnamefont
  {Preziosi}},\ }\href {\doibase 10.1103/RevModPhys.61.41} {\bibfield
  {journal} {\bibinfo  {journal} {Rev. Mod. Phys.}\ }\textbf {\bibinfo {volume}
  {61}},\ \bibinfo {pages} {41} (\bibinfo {year} {1989})}\BibitemShut {NoStop}%
\bibitem [{\citenamefont {Yu}, \citenamefont {Ouyang},\ and\ \citenamefont
  {Chen}(2021)}]{yu_perspective_2021}%
  \BibitemOpen
  \bibfield  {author} {\bibinfo {author} {\bibfnamefont {C.}~\bibnamefont
  {Yu}}, \bibinfo {author} {\bibfnamefont {Y.}~\bibnamefont {Ouyang}}, \ and\
  \bibinfo {author} {\bibfnamefont {J.}~\bibnamefont {Chen}},\ }\href {\doibase
  10.1063/5.0056315} {\bibfield  {journal} {\bibinfo  {journal} {J. Appl.
  Phys.}\ }\textbf {\bibinfo {volume} {130}},\ \bibinfo {pages} {010902}
  (\bibinfo {year} {2021})}\BibitemShut {NoStop}%
\bibitem [{\citenamefont {Li}\ \emph {et~al.}(2022)\citenamefont {Li},
  \citenamefont {Lee}, \citenamefont {Ou}, \citenamefont {Lee},\ and\
  \citenamefont {Shi}}]{LiShi_reexamination_hydrodynamic_JAP2022}%
  \BibitemOpen
  \bibfield  {author} {\bibinfo {author} {\bibfnamefont {X.}~\bibnamefont
  {Li}}, \bibinfo {author} {\bibfnamefont {H.}~\bibnamefont {Lee}}, \bibinfo
  {author} {\bibfnamefont {E.}~\bibnamefont {Ou}}, \bibinfo {author}
  {\bibfnamefont {S.}~\bibnamefont {Lee}}, \ and\ \bibinfo {author}
  {\bibfnamefont {L.}~\bibnamefont {Shi}},\ }\href {\doibase 10.1063/5.0078772}
  {\bibfield  {journal} {\bibinfo  {journal} {J. Appl. Phys.}\ }\textbf
  {\bibinfo {volume} {131}},\ \bibinfo {pages} {075104} (\bibinfo {year}
  {2022})}\BibitemShut {NoStop}%
\bibitem [{\citenamefont {Lee}\ \emph {et~al.}(2015)\citenamefont {Lee},
  \citenamefont {Broido}, \citenamefont {Esfarjani},\ and\ \citenamefont
  {Chen}}]{lee_hydrodynamic_2015}%
  \BibitemOpen
  \bibfield  {author} {\bibinfo {author} {\bibfnamefont {S.}~\bibnamefont
  {Lee}}, \bibinfo {author} {\bibfnamefont {D.}~\bibnamefont {Broido}},
  \bibinfo {author} {\bibfnamefont {K.}~\bibnamefont {Esfarjani}}, \ and\
  \bibinfo {author} {\bibfnamefont {G.}~\bibnamefont {Chen}},\ }\href {\doibase
  10.1038/ncomms7290} {\bibfield  {journal} {\bibinfo  {journal} {Nat.
  Commun.}\ }\textbf {\bibinfo {volume} {6}},\ \bibinfo {pages} {6290}
  (\bibinfo {year} {2015})}\BibitemShut {NoStop}%
\bibitem [{\citenamefont {Cepellotti}\ \emph {et~al.}(2015)\citenamefont
  {Cepellotti}, \citenamefont {Fugallo}, \citenamefont {Paulatto},
  \citenamefont {Lazzeri}, \citenamefont {Mauri},\ and\ \citenamefont
  {Marzari}}]{cepellotti_phonon_2015}%
  \BibitemOpen
  \bibfield  {author} {\bibinfo {author} {\bibfnamefont {A.}~\bibnamefont
  {Cepellotti}}, \bibinfo {author} {\bibfnamefont {G.}~\bibnamefont {Fugallo}},
  \bibinfo {author} {\bibfnamefont {L.}~\bibnamefont {Paulatto}}, \bibinfo
  {author} {\bibfnamefont {M.}~\bibnamefont {Lazzeri}}, \bibinfo {author}
  {\bibfnamefont {F.}~\bibnamefont {Mauri}}, \ and\ \bibinfo {author}
  {\bibfnamefont {N.}~\bibnamefont {Marzari}},\ }\href {\doibase
  10.1038/ncomms7400} {\bibfield  {journal} {\bibinfo  {journal} {Nat.
  Commun.}\ }\textbf {\bibinfo {volume} {6}},\ \bibinfo {pages} {6400}
  (\bibinfo {year} {2015})}\BibitemShut {NoStop}%
\bibitem [{\citenamefont {Lee}\ and\ \citenamefont {Lindsay}(2017)}]{lee2017}%
  \BibitemOpen
  \bibfield  {author} {\bibinfo {author} {\bibfnamefont {S.}~\bibnamefont
  {Lee}}\ and\ \bibinfo {author} {\bibfnamefont {L.}~\bibnamefont {Lindsay}},\
  }\href {\doibase 10.1103/PhysRevB.95.184304} {\bibfield  {journal} {\bibinfo
  {journal} {Phys. Rev. B}\ }\textbf {\bibinfo {volume} {95}},\ \bibinfo
  {pages} {184304} (\bibinfo {year} {2017})}\BibitemShut {NoStop}%
\bibitem [{\citenamefont {Luo}\ \emph {et~al.}(2019)\citenamefont {Luo},
  \citenamefont {Guo}, \citenamefont {Wang},\ and\ \citenamefont
  {Yi}}]{luo2019}%
  \BibitemOpen
  \bibfield  {author} {\bibinfo {author} {\bibfnamefont {X.-P.}\ \bibnamefont
  {Luo}}, \bibinfo {author} {\bibfnamefont {Y.-Y.}\ \bibnamefont {Guo}},
  \bibinfo {author} {\bibfnamefont {M.-R.}\ \bibnamefont {Wang}}, \ and\
  \bibinfo {author} {\bibfnamefont {H.-L.}\ \bibnamefont {Yi}},\ }\href
  {\doibase 10.1103/PhysRevB.100.155401} {\bibfield  {journal} {\bibinfo
  {journal} {Phys. Rev. B}\ }\textbf {\bibinfo {volume} {100}},\ \bibinfo
  {pages} {155401} (\bibinfo {year} {2019})}\BibitemShut {NoStop}%
\bibitem [{\citenamefont {Xu}(2021)}]{XU2021127402}%
  \BibitemOpen
  \bibfield  {author} {\bibinfo {author} {\bibfnamefont {M.}~\bibnamefont
  {Xu}},\ }\href {\doibase https://doi.org/10.1016/j.physleta.2021.127402}
  {\bibfield  {journal} {\bibinfo  {journal} {Phys. Lett. A}\ }\textbf
  {\bibinfo {volume} {404}},\ \bibinfo {pages} {127402} (\bibinfo {year}
  {2021})}\BibitemShut {NoStop}%
\bibitem [{\citenamefont {Shang}\ \emph {et~al.}(2022)\citenamefont {Shang},
  \citenamefont {Mao}, \citenamefont {Yang}, \citenamefont {Li},\ and\
  \citenamefont {L\"u}}]{PhysRevB.105.165423}%
  \BibitemOpen
  \bibfield  {author} {\bibinfo {author} {\bibfnamefont {M.-Y.}\ \bibnamefont
  {Shang}}, \bibinfo {author} {\bibfnamefont {W.-H.}\ \bibnamefont {Mao}},
  \bibinfo {author} {\bibfnamefont {N.}~\bibnamefont {Yang}}, \bibinfo {author}
  {\bibfnamefont {B.}~\bibnamefont {Li}}, \ and\ \bibinfo {author}
  {\bibfnamefont {J.-T.}\ \bibnamefont {L\"u}},\ }\href {\doibase
  10.1103/PhysRevB.105.165423} {\bibfield  {journal} {\bibinfo  {journal}
  {Phys. Rev. B}\ }\textbf {\bibinfo {volume} {105}},\ \bibinfo {pages}
  {165423} (\bibinfo {year} {2022})}\BibitemShut {NoStop}%
\bibitem [{\citenamefont {Shang}\ \emph {et~al.}(2020)\citenamefont {Shang},
  \citenamefont {Zhang}, \citenamefont {Guo},\ and\ \citenamefont
  {Lü}}]{shang_heat_2020}%
  \BibitemOpen
  \bibfield  {author} {\bibinfo {author} {\bibfnamefont {M.-Y.}\ \bibnamefont
  {Shang}}, \bibinfo {author} {\bibfnamefont {C.}~\bibnamefont {Zhang}},
  \bibinfo {author} {\bibfnamefont {Z.}~\bibnamefont {Guo}}, \ and\ \bibinfo
  {author} {\bibfnamefont {J.-T.}\ \bibnamefont {Lü}},\ }\href {\doibase
  10.1038/s41598-020-65221-8} {\bibfield  {journal} {\bibinfo  {journal} {Sci.
  Rep.}\ }\textbf {\bibinfo {volume} {10}},\ \bibinfo {pages} {8272} (\bibinfo
  {year} {2020})}\BibitemShut {NoStop}%
\bibitem [{\citenamefont {Benin}\ and\ \citenamefont
  {Maris}(1978)}]{Knudsenminimum_phonon78}%
  \BibitemOpen
  \bibfield  {author} {\bibinfo {author} {\bibfnamefont {D.}~\bibnamefont
  {Benin}}\ and\ \bibinfo {author} {\bibfnamefont {H.~J.}\ \bibnamefont
  {Maris}},\ }\href {\doibase 10.1103/PhysRevB.18.3112} {\bibfield  {journal}
  {\bibinfo  {journal} {Phys. Rev. B}\ }\textbf {\bibinfo {volume} {18}},\
  \bibinfo {pages} {3112} (\bibinfo {year} {1978})}\BibitemShut {NoStop}%
\bibitem [{\citenamefont {Li}\ and\ \citenamefont {Lee}(2018)}]{li2018a}%
  \BibitemOpen
  \bibfield  {author} {\bibinfo {author} {\bibfnamefont {X.}~\bibnamefont
  {Li}}\ and\ \bibinfo {author} {\bibfnamefont {S.}~\bibnamefont {Lee}},\
  }\href {\doibase 10.1103/PhysRevB.97.094309} {\bibfield  {journal} {\bibinfo
  {journal} {Phys. Rev. B}\ }\textbf {\bibinfo {volume} {97}},\ \bibinfo
  {pages} {094309} (\bibinfo {year} {2018})}\BibitemShut {NoStop}%
\bibitem [{\citenamefont {Guo}\ and\ \citenamefont
  {Wang}(2017)}]{wangmr17callaway}%
  \BibitemOpen
  \bibfield  {author} {\bibinfo {author} {\bibfnamefont {Y.}~\bibnamefont
  {Guo}}\ and\ \bibinfo {author} {\bibfnamefont {M.}~\bibnamefont {Wang}},\
  }\href {\doibase 10.1103/PhysRevB.96.134312} {\bibfield  {journal} {\bibinfo
  {journal} {Phys. Rev. B}\ }\textbf {\bibinfo {volume} {96}},\ \bibinfo
  {pages} {134312} (\bibinfo {year} {2017})}\BibitemShut {NoStop}%
\bibitem [{\citenamefont {Huberman}\ \emph {et~al.}(2019)\citenamefont
  {Huberman}, \citenamefont {Duncan}, \citenamefont {Chen}, \citenamefont
  {Song}, \citenamefont {Chiloyan}, \citenamefont {Ding}, \citenamefont
  {Maznev}, \citenamefont {Chen},\ and\ \citenamefont
  {Nelson}}]{huberman_observation_2019}%
  \BibitemOpen
  \bibfield  {author} {\bibinfo {author} {\bibfnamefont {S.}~\bibnamefont
  {Huberman}}, \bibinfo {author} {\bibfnamefont {R.~A.}\ \bibnamefont
  {Duncan}}, \bibinfo {author} {\bibfnamefont {K.}~\bibnamefont {Chen}},
  \bibinfo {author} {\bibfnamefont {B.}~\bibnamefont {Song}}, \bibinfo {author}
  {\bibfnamefont {V.}~\bibnamefont {Chiloyan}}, \bibinfo {author}
  {\bibfnamefont {Z.}~\bibnamefont {Ding}}, \bibinfo {author} {\bibfnamefont
  {A.~A.}\ \bibnamefont {Maznev}}, \bibinfo {author} {\bibfnamefont
  {G.}~\bibnamefont {Chen}}, \ and\ \bibinfo {author} {\bibfnamefont {K.~A.}\
  \bibnamefont {Nelson}},\ }\href {\doibase 10.1126/science.aav3548} {\bibfield
   {journal} {\bibinfo  {journal} {Science}\ }\textbf {\bibinfo {volume}
  {364}},\ \bibinfo {pages} {375} (\bibinfo {year} {2019})}\BibitemShut
  {NoStop}%
\bibitem [{\citenamefont {Ding}\ \emph {et~al.}(2022)\citenamefont {Ding},
  \citenamefont {Chen}, \citenamefont {Song}, \citenamefont {Shin},
  \citenamefont {Maznev}, \citenamefont {Nelson},\ and\ \citenamefont
  {Chen}}]{ding_observation_2022}%
  \BibitemOpen
  \bibfield  {author} {\bibinfo {author} {\bibfnamefont {Z.}~\bibnamefont
  {Ding}}, \bibinfo {author} {\bibfnamefont {K.}~\bibnamefont {Chen}}, \bibinfo
  {author} {\bibfnamefont {B.}~\bibnamefont {Song}}, \bibinfo {author}
  {\bibfnamefont {J.}~\bibnamefont {Shin}}, \bibinfo {author} {\bibfnamefont
  {A.~A.}\ \bibnamefont {Maznev}}, \bibinfo {author} {\bibfnamefont {K.~A.}\
  \bibnamefont {Nelson}}, \ and\ \bibinfo {author} {\bibfnamefont
  {G.}~\bibnamefont {Chen}},\ }\href {\doibase 10.1038/s41467-021-27907-z}
  {\bibfield  {journal} {\bibinfo  {journal} {Nat. Commun.}\ }\textbf {\bibinfo
  {volume} {13}},\ \bibinfo {pages} {285} (\bibinfo {year} {2022})}\BibitemShut
  {NoStop}%
\bibitem [{\citenamefont {Martelli}\ \emph {et~al.}(2018)\citenamefont
  {Martelli}, \citenamefont {Jim\'enez}, \citenamefont {Continentino},
  \citenamefont {Baggio-Saitovitch},\ and\ \citenamefont
  {Behnia}}]{PhysRevLett_Strontium_Titanate}%
  \BibitemOpen
  \bibfield  {author} {\bibinfo {author} {\bibfnamefont {V.}~\bibnamefont
  {Martelli}}, \bibinfo {author} {\bibfnamefont {J.~L.}\ \bibnamefont
  {Jim\'enez}}, \bibinfo {author} {\bibfnamefont {M.}~\bibnamefont
  {Continentino}}, \bibinfo {author} {\bibfnamefont {E.}~\bibnamefont
  {Baggio-Saitovitch}}, \ and\ \bibinfo {author} {\bibfnamefont
  {K.}~\bibnamefont {Behnia}},\ }\href {\doibase
  10.1103/PhysRevLett.120.125901} {\bibfield  {journal} {\bibinfo  {journal}
  {Phys. Rev. Lett.}\ }\textbf {\bibinfo {volume} {120}},\ \bibinfo {pages}
  {125901} (\bibinfo {year} {2018})}\BibitemShut {NoStop}%
\bibitem [{\citenamefont {Machida}\ \emph {et~al.}(2020)\citenamefont
  {Machida}, \citenamefont {Matsumoto}, \citenamefont {Isono},\ and\
  \citenamefont {Behnia}}]{machida2020}%
  \BibitemOpen
  \bibfield  {author} {\bibinfo {author} {\bibfnamefont {Y.}~\bibnamefont
  {Machida}}, \bibinfo {author} {\bibfnamefont {N.}~\bibnamefont {Matsumoto}},
  \bibinfo {author} {\bibfnamefont {T.}~\bibnamefont {Isono}}, \ and\ \bibinfo
  {author} {\bibfnamefont {K.}~\bibnamefont {Behnia}},\ }\href {\doibase
  10.1126/science.aaz8043} {\bibfield  {journal} {\bibinfo  {journal}
  {Science}\ }\textbf {\bibinfo {volume} {367}},\ \bibinfo {pages} {309}
  (\bibinfo {year} {2020})}\BibitemShut {NoStop}%
\bibitem [{\citenamefont {Machida}\ \emph {et~al.}(2018)\citenamefont
  {Machida}, \citenamefont {Subedi}, \citenamefont {Akiba}, \citenamefont
  {Miyake}, \citenamefont {Tokunaga}, \citenamefont {Akahama}, \citenamefont
  {Izawa},\ and\ \citenamefont {Behnia}}]{machida2018}%
  \BibitemOpen
  \bibfield  {author} {\bibinfo {author} {\bibfnamefont {Y.}~\bibnamefont
  {Machida}}, \bibinfo {author} {\bibfnamefont {A.}~\bibnamefont {Subedi}},
  \bibinfo {author} {\bibfnamefont {K.}~\bibnamefont {Akiba}}, \bibinfo
  {author} {\bibfnamefont {A.}~\bibnamefont {Miyake}}, \bibinfo {author}
  {\bibfnamefont {M.}~\bibnamefont {Tokunaga}}, \bibinfo {author}
  {\bibfnamefont {Y.}~\bibnamefont {Akahama}}, \bibinfo {author} {\bibfnamefont
  {K.}~\bibnamefont {Izawa}}, \ and\ \bibinfo {author} {\bibfnamefont
  {K.}~\bibnamefont {Behnia}},\ }\href {\doibase 10.1126/sciadv.aat3374}
  {\bibfield  {journal} {\bibinfo  {journal} {Sci. Adv.}\ }\textbf {\bibinfo
  {volume} {4}},\ \bibinfo {pages} {eaat3374} (\bibinfo {year}
  {2018})}\BibitemShut {NoStop}%
\bibitem [{\citenamefont {Huang}\ \emph {et~al.}(2023)\citenamefont {Huang},
  \citenamefont {Guo}, \citenamefont {Wu}, \citenamefont {Masubuchi},
  \citenamefont {Watanabe}, \citenamefont {Taniguchi}, \citenamefont {Zhang},
  \citenamefont {Volz}, \citenamefont {Machida},\ and\ \citenamefont
  {Nomura}}]{huang2023observation}%
  \BibitemOpen
  \bibfield  {author} {\bibinfo {author} {\bibfnamefont {X.}~\bibnamefont
  {Huang}}, \bibinfo {author} {\bibfnamefont {Y.}~\bibnamefont {Guo}}, \bibinfo
  {author} {\bibfnamefont {Y.}~\bibnamefont {Wu}}, \bibinfo {author}
  {\bibfnamefont {S.}~\bibnamefont {Masubuchi}}, \bibinfo {author}
  {\bibfnamefont {K.}~\bibnamefont {Watanabe}}, \bibinfo {author}
  {\bibfnamefont {T.}~\bibnamefont {Taniguchi}}, \bibinfo {author}
  {\bibfnamefont {Z.}~\bibnamefont {Zhang}}, \bibinfo {author} {\bibfnamefont
  {S.}~\bibnamefont {Volz}}, \bibinfo {author} {\bibfnamefont {T.}~\bibnamefont
  {Machida}}, \ and\ \bibinfo {author} {\bibfnamefont {M.}~\bibnamefont
  {Nomura}},\ }\href@noop {} {\bibfield  {journal} {\bibinfo  {journal} {Nat.
  Commun.}\ }\textbf {\bibinfo {volume} {14}},\ \bibinfo {pages} {2044}
  (\bibinfo {year} {2023})}\BibitemShut {NoStop}%
\bibitem [{\citenamefont {Beardo}\ \emph {et~al.}(2020)\citenamefont {Beardo},
  \citenamefont {Hennessy}, \citenamefont {Sendra}, \citenamefont {Camacho},
  \citenamefont {Myers}, \citenamefont {Bafaluy},\ and\ \citenamefont
  {Alvarez}}]{PhysRevB.101.075303}%
  \BibitemOpen
  \bibfield  {author} {\bibinfo {author} {\bibfnamefont {A.}~\bibnamefont
  {Beardo}}, \bibinfo {author} {\bibfnamefont {M.~G.}\ \bibnamefont
  {Hennessy}}, \bibinfo {author} {\bibfnamefont {L.}~\bibnamefont {Sendra}},
  \bibinfo {author} {\bibfnamefont {J.}~\bibnamefont {Camacho}}, \bibinfo
  {author} {\bibfnamefont {T.~G.}\ \bibnamefont {Myers}}, \bibinfo {author}
  {\bibfnamefont {J.}~\bibnamefont {Bafaluy}}, \ and\ \bibinfo {author}
  {\bibfnamefont {F.~X.}\ \bibnamefont {Alvarez}},\ }\href {\doibase
  10.1103/PhysRevB.101.075303} {\bibfield  {journal} {\bibinfo  {journal}
  {Phys. Rev. B}\ }\textbf {\bibinfo {volume} {101}},\ \bibinfo {pages}
  {075303} (\bibinfo {year} {2020})}\BibitemShut {NoStop}%
\bibitem [{\citenamefont {Beardo}\ \emph {et~al.}(2021)\citenamefont {Beardo},
  \citenamefont {López-Suárez}, \citenamefont {Pérez}, \citenamefont
  {Sendra}, \citenamefont {Alonso}, \citenamefont {Melis}, \citenamefont
  {Bafaluy}, \citenamefont {Camacho}, \citenamefont {Colombo}, \citenamefont
  {Rurali}, \citenamefont {Alvarez},\ and\ \citenamefont
  {Reparaz}}]{beardo_observation_2021}%
  \BibitemOpen
  \bibfield  {author} {\bibinfo {author} {\bibfnamefont {A.}~\bibnamefont
  {Beardo}}, \bibinfo {author} {\bibfnamefont {M.}~\bibnamefont
  {López-Suárez}}, \bibinfo {author} {\bibfnamefont {L.~A.}\ \bibnamefont
  {Pérez}}, \bibinfo {author} {\bibfnamefont {L.}~\bibnamefont {Sendra}},
  \bibinfo {author} {\bibfnamefont {M.~I.}\ \bibnamefont {Alonso}}, \bibinfo
  {author} {\bibfnamefont {C.}~\bibnamefont {Melis}}, \bibinfo {author}
  {\bibfnamefont {J.}~\bibnamefont {Bafaluy}}, \bibinfo {author} {\bibfnamefont
  {J.}~\bibnamefont {Camacho}}, \bibinfo {author} {\bibfnamefont
  {L.}~\bibnamefont {Colombo}}, \bibinfo {author} {\bibfnamefont
  {R.}~\bibnamefont {Rurali}}, \bibinfo {author} {\bibfnamefont {F.~X.}\
  \bibnamefont {Alvarez}}, \ and\ \bibinfo {author} {\bibfnamefont {J.~S.}\
  \bibnamefont {Reparaz}},\ }\href {\doibase 10.1126/sciadv.abg4677} {\bibfield
   {journal} {\bibinfo  {journal} {Sci. Adv.}\ }\textbf {\bibinfo {volume}
  {7}},\ \bibinfo {pages} {eabg4677} (\bibinfo {year} {2021})}\BibitemShut
  {NoStop}%
\bibitem [{\citenamefont {Chiloyan}\ \emph {et~al.}(2021)\citenamefont
  {Chiloyan}, \citenamefont {Huberman}, \citenamefont {Ding}, \citenamefont
  {Mendoza}, \citenamefont {Maznev}, \citenamefont {Nelson},\ and\
  \citenamefont {Chen}}]{PhysRevB.104.245424}%
  \BibitemOpen
  \bibfield  {author} {\bibinfo {author} {\bibfnamefont {V.}~\bibnamefont
  {Chiloyan}}, \bibinfo {author} {\bibfnamefont {S.}~\bibnamefont {Huberman}},
  \bibinfo {author} {\bibfnamefont {Z.}~\bibnamefont {Ding}}, \bibinfo {author}
  {\bibfnamefont {J.}~\bibnamefont {Mendoza}}, \bibinfo {author} {\bibfnamefont
  {A.~A.}\ \bibnamefont {Maznev}}, \bibinfo {author} {\bibfnamefont {K.~A.}\
  \bibnamefont {Nelson}}, \ and\ \bibinfo {author} {\bibfnamefont
  {G.}~\bibnamefont {Chen}},\ }\href {\doibase 10.1103/PhysRevB.104.245424}
  {\bibfield  {journal} {\bibinfo  {journal} {Phys. Rev. B}\ }\textbf {\bibinfo
  {volume} {104}},\ \bibinfo {pages} {245424} (\bibinfo {year}
  {2021})}\BibitemShut {NoStop}%
\bibitem [{\citenamefont {Zhang}, \citenamefont {Huberman},\ and\ \citenamefont
  {Wu}(2022)}]{heatwaves_2022chuang}%
  \BibitemOpen
  \bibfield  {author} {\bibinfo {author} {\bibfnamefont {C.}~\bibnamefont
  {Zhang}}, \bibinfo {author} {\bibfnamefont {S.}~\bibnamefont {Huberman}}, \
  and\ \bibinfo {author} {\bibfnamefont {L.}~\bibnamefont {Wu}},\ }\href
  {\doibase 10.1063/5.0102227} {\bibfield  {journal} {\bibinfo  {journal} {J.
  Appl. Phys.}\ }\textbf {\bibinfo {volume} {132}},\ \bibinfo {pages} {085103}
  (\bibinfo {year} {2022})}\BibitemShut {NoStop}%
\bibitem [{\citenamefont {Hardy}(1970)}]{PhysRevB_SECOND_SOUND}%
  \BibitemOpen
  \bibfield  {author} {\bibinfo {author} {\bibfnamefont {R.~J.}\ \bibnamefont
  {Hardy}},\ }\href {\doibase 10.1103/PhysRevB.2.1193} {\bibfield  {journal}
  {\bibinfo  {journal} {Phys. Rev. B}\ }\textbf {\bibinfo {volume} {2}},\
  \bibinfo {pages} {1193} (\bibinfo {year} {1970})}\BibitemShut {NoStop}%
\bibitem [{\citenamefont {Zhang}, \citenamefont {Guo},\ and\ \citenamefont
  {Chen}(2019)}]{ZHANG20191366}%
  \BibitemOpen
  \bibfield  {author} {\bibinfo {author} {\bibfnamefont {C.}~\bibnamefont
  {Zhang}}, \bibinfo {author} {\bibfnamefont {Z.}~\bibnamefont {Guo}}, \ and\
  \bibinfo {author} {\bibfnamefont {S.}~\bibnamefont {Chen}},\ }\href {\doibase
  10.1016/j.ijheatmasstransfer.2018.10.141} {\bibfield  {journal} {\bibinfo
  {journal} {Int. J. Heat Mass Transfer}\ }\textbf {\bibinfo {volume} {130}},\
  \bibinfo {pages} {1366} (\bibinfo {year} {2019})}\BibitemShut {NoStop}%
\bibitem [{\citenamefont {Zhang}, \citenamefont {Chen},\ and\ \citenamefont
  {Guo}(2021)}]{zhang_heat_2021}%
  \BibitemOpen
  \bibfield  {author} {\bibinfo {author} {\bibfnamefont {C.}~\bibnamefont
  {Zhang}}, \bibinfo {author} {\bibfnamefont {S.}~\bibnamefont {Chen}}, \ and\
  \bibinfo {author} {\bibfnamefont {Z.}~\bibnamefont {Guo}},\ }\href {\doibase
  10.1016/j.ijheatmasstransfer.2021.121282} {\bibfield  {journal} {\bibinfo
  {journal} {Int. J. Heat Mass Transfer}\ }\textbf {\bibinfo {volume} {176}},\
  \bibinfo {pages} {121282} (\bibinfo {year} {2021})}\BibitemShut {NoStop}%
\bibitem [{\citenamefont {Beardo}\ \emph {et~al.}(2019)\citenamefont {Beardo},
  \citenamefont {Calvo-Schwarzw\"alder}, \citenamefont {Camacho}, \citenamefont
  {Myers}, \citenamefont {Torres}, \citenamefont {Sendra}, \citenamefont
  {Alvarez},\ and\ \citenamefont {Bafaluy}}]{PhysRevApplied.11.034003}%
  \BibitemOpen
  \bibfield  {author} {\bibinfo {author} {\bibfnamefont {A.}~\bibnamefont
  {Beardo}}, \bibinfo {author} {\bibfnamefont {M.}~\bibnamefont
  {Calvo-Schwarzw\"alder}}, \bibinfo {author} {\bibfnamefont {J.}~\bibnamefont
  {Camacho}}, \bibinfo {author} {\bibfnamefont {T.}~\bibnamefont {Myers}},
  \bibinfo {author} {\bibfnamefont {P.}~\bibnamefont {Torres}}, \bibinfo
  {author} {\bibfnamefont {L.}~\bibnamefont {Sendra}}, \bibinfo {author}
  {\bibfnamefont {F.}~\bibnamefont {Alvarez}}, \ and\ \bibinfo {author}
  {\bibfnamefont {J.}~\bibnamefont {Bafaluy}},\ }\href {\doibase
  10.1103/PhysRevApplied.11.034003} {\bibfield  {journal} {\bibinfo  {journal}
  {Phys. Rev. Applied}\ }\textbf {\bibinfo {volume} {11}},\ \bibinfo {pages}
  {034003} (\bibinfo {year} {2019})}\BibitemShut {NoStop}%
\bibitem [{\citenamefont {Zhang}\ \emph {et~al.}(2022)\citenamefont {Zhang},
  \citenamefont {Ma}, \citenamefont {Shang}, \citenamefont {Wan}, \citenamefont
  {Lü}, \citenamefont {Guo}, \citenamefont {Li},\ and\ \citenamefont
  {Yang}}]{chuang2021graded}%
  \BibitemOpen
  \bibfield  {author} {\bibinfo {author} {\bibfnamefont {C.}~\bibnamefont
  {Zhang}}, \bibinfo {author} {\bibfnamefont {D.}~\bibnamefont {Ma}}, \bibinfo
  {author} {\bibfnamefont {M.}~\bibnamefont {Shang}}, \bibinfo {author}
  {\bibfnamefont {X.}~\bibnamefont {Wan}}, \bibinfo {author} {\bibfnamefont
  {J.-T.}\ \bibnamefont {Lü}}, \bibinfo {author} {\bibfnamefont
  {Z.}~\bibnamefont {Guo}}, \bibinfo {author} {\bibfnamefont {B.}~\bibnamefont
  {Li}}, \ and\ \bibinfo {author} {\bibfnamefont {N.}~\bibnamefont {Yang}},\
  }\href {\doibase https://doi.org/10.1016/j.mtphys.2022.100605} {\bibfield
  {journal} {\bibinfo  {journal} {Mater. Today Phys.}\ }\textbf {\bibinfo
  {volume} {22}},\ \bibinfo {pages} {100605} (\bibinfo {year}
  {2022})}\BibitemShut {NoStop}%
\bibitem [{\citenamefont {Zhang}\ and\ \citenamefont
  {Guo}(2021)}]{zhang_transient_2021}%
  \BibitemOpen
  \bibfield  {author} {\bibinfo {author} {\bibfnamefont {C.}~\bibnamefont
  {Zhang}}\ and\ \bibinfo {author} {\bibfnamefont {Z.}~\bibnamefont {Guo}},\
  }\href {\doibase 10.1016/j.ijheatmasstransfer.2021.121847} {\bibfield
  {journal} {\bibinfo  {journal} {Int. J. Heat Mass Transfer}\ }\textbf
  {\bibinfo {volume} {181}},\ \bibinfo {pages} {121847} (\bibinfo {year}
  {2021})}\BibitemShut {NoStop}%
\bibitem [{\citenamefont {Jeong}\ \emph {et~al.}(2021)\citenamefont {Jeong},
  \citenamefont {Li}, \citenamefont {Lee}, \citenamefont {Shi},\ and\
  \citenamefont {Wang}}]{PhysRevLett.127.085901}%
  \BibitemOpen
  \bibfield  {author} {\bibinfo {author} {\bibfnamefont {J.}~\bibnamefont
  {Jeong}}, \bibinfo {author} {\bibfnamefont {X.}~\bibnamefont {Li}}, \bibinfo
  {author} {\bibfnamefont {S.}~\bibnamefont {Lee}}, \bibinfo {author}
  {\bibfnamefont {L.}~\bibnamefont {Shi}}, \ and\ \bibinfo {author}
  {\bibfnamefont {Y.}~\bibnamefont {Wang}},\ }\href {\doibase
  10.1103/PhysRevLett.127.085901} {\bibfield  {journal} {\bibinfo  {journal}
  {Phys. Rev. Lett.}\ }\textbf {\bibinfo {volume} {127}},\ \bibinfo {pages}
  {085901} (\bibinfo {year} {2021})}\BibitemShut {NoStop}%
\bibitem [{\citenamefont {Xu}\ \emph {et~al.}(2020)\citenamefont {Xu},
  \citenamefont {Fan}, \citenamefont {Wang}, \citenamefont {Zhang},\ and\
  \citenamefont {Wang}}]{xu_raman-based_2020}%
  \BibitemOpen
  \bibfield  {author} {\bibinfo {author} {\bibfnamefont {S.}~\bibnamefont
  {Xu}}, \bibinfo {author} {\bibfnamefont {A.}~\bibnamefont {Fan}}, \bibinfo
  {author} {\bibfnamefont {H.}~\bibnamefont {Wang}}, \bibinfo {author}
  {\bibfnamefont {X.}~\bibnamefont {Zhang}}, \ and\ \bibinfo {author}
  {\bibfnamefont {X.}~\bibnamefont {Wang}},\ }\href {\doibase
  10.1016/j.ijheatmasstransfer.2020.119751} {\bibfield  {journal} {\bibinfo
  {journal} {Int. J. Heat Mass Transfer}\ }\textbf {\bibinfo {volume} {154}},\
  \bibinfo {pages} {119751} (\bibinfo {year} {2020})}\BibitemShut {NoStop}%
\bibitem [{\citenamefont {Fan}\ \emph {et~al.}(2019)\citenamefont {Fan},
  \citenamefont {Hu}, \citenamefont {Wang}, \citenamefont {Ma},\ and\
  \citenamefont {Zhang}}]{fan_dual-wavelength_2019}%
  \BibitemOpen
  \bibfield  {author} {\bibinfo {author} {\bibfnamefont {A.}~\bibnamefont
  {Fan}}, \bibinfo {author} {\bibfnamefont {Y.}~\bibnamefont {Hu}}, \bibinfo
  {author} {\bibfnamefont {H.}~\bibnamefont {Wang}}, \bibinfo {author}
  {\bibfnamefont {W.}~\bibnamefont {Ma}}, \ and\ \bibinfo {author}
  {\bibfnamefont {X.}~\bibnamefont {Zhang}},\ }\href {\doibase
  10.1016/j.ijheatmasstransfer.2019.118460} {\bibfield  {journal} {\bibinfo
  {journal} {Int. J. Heat Mass Transfer}\ }\textbf {\bibinfo {volume} {143}},\
  \bibinfo {pages} {118460} (\bibinfo {year} {2019})}\BibitemShut {NoStop}%
\bibitem [{\citenamefont {Callaway}(1959)}]{PhysRev_callaway}%
  \BibitemOpen
  \bibfield  {author} {\bibinfo {author} {\bibfnamefont {J.}~\bibnamefont
  {Callaway}},\ }\href {\doibase 10.1103/PhysRev.113.1046} {\bibfield
  {journal} {\bibinfo  {journal} {Phys. Rev.}\ }\textbf {\bibinfo {volume}
  {113}},\ \bibinfo {pages} {1046} (\bibinfo {year} {1959})}\BibitemShut
  {NoStop}%
\bibitem [{\citenamefont {Ding}\ \emph {et~al.}(2018)\citenamefont {Ding},
  \citenamefont {Zhou}, \citenamefont {Song}, \citenamefont {Chiloyan},
  \citenamefont {Li}, \citenamefont {Liu},\ and\ \citenamefont
  {Chen}}]{nanoletterchengang_2018}%
  \BibitemOpen
  \bibfield  {author} {\bibinfo {author} {\bibfnamefont {Z.}~\bibnamefont
  {Ding}}, \bibinfo {author} {\bibfnamefont {J.}~\bibnamefont {Zhou}}, \bibinfo
  {author} {\bibfnamefont {B.}~\bibnamefont {Song}}, \bibinfo {author}
  {\bibfnamefont {V.}~\bibnamefont {Chiloyan}}, \bibinfo {author}
  {\bibfnamefont {M.}~\bibnamefont {Li}}, \bibinfo {author} {\bibfnamefont
  {T.-H.}\ \bibnamefont {Liu}}, \ and\ \bibinfo {author} {\bibfnamefont
  {G.}~\bibnamefont {Chen}},\ }\href {\doibase 10.1021/acs.nanolett.7b04932}
  {\bibfield  {journal} {\bibinfo  {journal} {Nano Lett.}\ }\textbf {\bibinfo
  {volume} {18}},\ \bibinfo {pages} {638} (\bibinfo {year} {2018})}\BibitemShut
  {NoStop}%
\bibitem [{\citenamefont {Nie}\ and\ \citenamefont
  {Cao}(2020)}]{nie2020thermal}%
  \BibitemOpen
  \bibfield  {author} {\bibinfo {author} {\bibfnamefont {B.-D.}\ \bibnamefont
  {Nie}}\ and\ \bibinfo {author} {\bibfnamefont {B.-Y.}\ \bibnamefont {Cao}},\
  }\href {\doibase 10.1080/15567265.2020.1755399} {\bibfield  {journal}
  {\bibinfo  {journal} {Nanosc. Microsc. Therm.}\ }\textbf {\bibinfo {volume}
  {24}},\ \bibinfo {pages} {94} (\bibinfo {year} {2020})}\BibitemShut {NoStop}%
\bibitem [{\citenamefont {Chiloyan}\ \emph {et~al.}(2020)\citenamefont
  {Chiloyan}, \citenamefont {Huberman}, \citenamefont {Maznev}, \citenamefont
  {Nelson},\ and\ \citenamefont {Chen}}]{APLnonthermal2020}%
  \BibitemOpen
  \bibfield  {author} {\bibinfo {author} {\bibfnamefont {V.}~\bibnamefont
  {Chiloyan}}, \bibinfo {author} {\bibfnamefont {S.}~\bibnamefont {Huberman}},
  \bibinfo {author} {\bibfnamefont {A.~A.}\ \bibnamefont {Maznev}}, \bibinfo
  {author} {\bibfnamefont {K.~A.}\ \bibnamefont {Nelson}}, \ and\ \bibinfo
  {author} {\bibfnamefont {G.}~\bibnamefont {Chen}},\ }\href {\doibase
  10.1063/1.5139069} {\bibfield  {journal} {\bibinfo  {journal} {Appl. Phys.
  Lett.}\ }\textbf {\bibinfo {volume} {116}},\ \bibinfo {pages} {163102}
  (\bibinfo {year} {2020})}\BibitemShut {NoStop}%
\bibitem [{\citenamefont {Hua}\ and\ \citenamefont
  {Minnich}(2018)}]{PhysRevB.97.014307}%
  \BibitemOpen
  \bibfield  {author} {\bibinfo {author} {\bibfnamefont {C.}~\bibnamefont
  {Hua}}\ and\ \bibinfo {author} {\bibfnamefont {A.~J.}\ \bibnamefont
  {Minnich}},\ }\href {\doibase 10.1103/PhysRevB.97.014307} {\bibfield
  {journal} {\bibinfo  {journal} {Phys. Rev. B}\ }\textbf {\bibinfo {volume}
  {97}},\ \bibinfo {pages} {014307} (\bibinfo {year} {2018})}\BibitemShut
  {NoStop}%
\bibitem [{\citenamefont {Vallabhaneni}\ \emph {et~al.}(2016)\citenamefont
  {Vallabhaneni}, \citenamefont {Singh}, \citenamefont {Bao}, \citenamefont
  {Murthy},\ and\ \citenamefont {Ruan}}]{PhysRevB.93.125432}%
  \BibitemOpen
  \bibfield  {author} {\bibinfo {author} {\bibfnamefont {A.~K.}\ \bibnamefont
  {Vallabhaneni}}, \bibinfo {author} {\bibfnamefont {D.}~\bibnamefont {Singh}},
  \bibinfo {author} {\bibfnamefont {H.}~\bibnamefont {Bao}}, \bibinfo {author}
  {\bibfnamefont {J.}~\bibnamefont {Murthy}}, \ and\ \bibinfo {author}
  {\bibfnamefont {X.}~\bibnamefont {Ruan}},\ }\href {\doibase
  10.1103/PhysRevB.93.125432} {\bibfield  {journal} {\bibinfo  {journal} {Phys.
  Rev. B}\ }\textbf {\bibinfo {volume} {93}},\ \bibinfo {pages} {125432}
  (\bibinfo {year} {2016})}\BibitemShut {NoStop}%
\bibitem [{\citenamefont {Guo}\ and\ \citenamefont
  {Xu}(2021)}]{guo_progress_DUGKS}%
  \BibitemOpen
  \bibfield  {author} {\bibinfo {author} {\bibfnamefont {Z.}~\bibnamefont
  {Guo}}\ and\ \bibinfo {author} {\bibfnamefont {K.}~\bibnamefont {Xu}},\
  }\href {\doibase 10.1186/s42774-020-00058-3} {\bibfield  {journal} {\bibinfo
  {journal} {Adva. Aerodyn.}\ }\textbf {\bibinfo {volume} {3}},\ \bibinfo
  {pages} {6} (\bibinfo {year} {2021})}\BibitemShut {NoStop}%
\bibitem [{\citenamefont {Simoncelli}, \citenamefont {Marzari},\ and\
  \citenamefont {Cepellotti}(2020)}]{PhysRevX.10.011019}%
  \BibitemOpen
  \bibfield  {author} {\bibinfo {author} {\bibfnamefont {M.}~\bibnamefont
  {Simoncelli}}, \bibinfo {author} {\bibfnamefont {N.}~\bibnamefont {Marzari}},
  \ and\ \bibinfo {author} {\bibfnamefont {A.}~\bibnamefont {Cepellotti}},\
  }\href {\doibase 10.1103/PhysRevX.10.011019} {\bibfield  {journal} {\bibinfo
  {journal} {Phys. Rev. X}\ }\textbf {\bibinfo {volume} {10}},\ \bibinfo
  {pages} {011019} (\bibinfo {year} {2020})}\BibitemShut {NoStop}%
\bibitem [{\citenamefont {Gu}\ \emph {et~al.}(2018)\citenamefont {Gu},
  \citenamefont {Wei}, \citenamefont {Yin}, \citenamefont {Li},\ and\
  \citenamefont {Yang}}]{RevModPhys.90.041002}%
  \BibitemOpen
  \bibfield  {author} {\bibinfo {author} {\bibfnamefont {X.}~\bibnamefont
  {Gu}}, \bibinfo {author} {\bibfnamefont {Y.}~\bibnamefont {Wei}}, \bibinfo
  {author} {\bibfnamefont {X.}~\bibnamefont {Yin}}, \bibinfo {author}
  {\bibfnamefont {B.}~\bibnamefont {Li}}, \ and\ \bibinfo {author}
  {\bibfnamefont {R.}~\bibnamefont {Yang}},\ }\href {\doibase
  10.1103/RevModPhys.90.041002} {\bibfield  {journal} {\bibinfo  {journal}
  {Rev. Mod. Phys.}\ }\textbf {\bibinfo {volume} {90}},\ \bibinfo {pages}
  {041002} (\bibinfo {year} {2018})}\BibitemShut {NoStop}%
\bibitem [{\citenamefont {Deng}\ \emph {et~al.}(2021)\citenamefont {Deng},
  \citenamefont {Huang}, \citenamefont {An},\ and\ \citenamefont
  {Yang}}]{MTP_review_nuo_2021}%
  \BibitemOpen
  \bibfield  {author} {\bibinfo {author} {\bibfnamefont {C.}~\bibnamefont
  {Deng}}, \bibinfo {author} {\bibfnamefont {Y.}~\bibnamefont {Huang}},
  \bibinfo {author} {\bibfnamefont {M.}~\bibnamefont {An}}, \ and\ \bibinfo
  {author} {\bibfnamefont {N.}~\bibnamefont {Yang}},\ }\href {\doibase
  https://doi.org/10.1016/j.mtphys.2020.100305} {\bibfield  {journal} {\bibinfo
   {journal} {Mater. Today Phys.}\ }\textbf {\bibinfo {volume} {16}},\ \bibinfo
  {pages} {100305} (\bibinfo {year} {2021})}\BibitemShut {NoStop}%
\bibitem [{\citenamefont {Chen}(2005)}]{ChenG05Oxford}%
  \BibitemOpen
  \bibfield  {author} {\bibinfo {author} {\bibfnamefont {G.}~\bibnamefont
  {Chen}},\ }\href
  {https://global.oup.com/ushe/product/nanoscale-energy-transport-and-conversion-9780195159424?cc=cn&lang=en&}
  {\emph {\bibinfo {title} {Nanoscale energy transport and conversion: {A}
  parallel treatment of electrons, molecules, phonons, and photons}}}\
  (\bibinfo  {publisher} {Oxford University Press},\ \bibinfo {year}
  {2005})\BibitemShut {NoStop}%
\bibitem [{\citenamefont {Kubo~R.}\ and\ \citenamefont
  {N.}(1991)}]{Kubo1991statistical}%
  \BibitemOpen
  \bibfield  {author} {\bibinfo {author} {\bibfnamefont {T.~M.}\ \bibnamefont
  {Kubo~R.}}\ and\ \bibinfo {author} {\bibfnamefont {H.}~\bibnamefont {N.}},\
  }\href {https://doi.org/10.1007/978-3-642-58244-8} {\emph {\bibinfo {title}
  {Statistical Physics II Nonequilibrium Statistical Mechanics}}},\ Springer
  Series in Solid State Sciences\ (\bibinfo  {publisher} {Springer, Berlin,
  Heidelberg},\ \bibinfo {year} {1991})\BibitemShut {NoStop}%
\end{thebibliography}%

\end{document}